\documentstyle[sprocl,epsfig]{article}

\begin{document}

\title{Instabilities in a Mean-field dynamics of Asymmetric Nuclear Matter}

\author{V. Greco, V. Baran, M. Colonna, M. Di Toro}

\address{Laboratorio Nazionale del Sud, Via S. Sofia 44,
I-95123 Catania, Italy and Universit\`a degli Studi di Catania\\
E-mail: greco,baran,ditoro,colonna@lns.infn.it}

\author{G. Fabbri, F. Matera}

\address{Istituto Nazionale di Fisica Nucleare, Sezione di Firenze and\\
Dipartimento di Fisica, L.go E.Fermi 2, I-50125, Firenze, Italy \\
E-mail: matera,gfabbri@fi.infn.it}  

\maketitle

\abstracts{We discuss the features of instabilities in asymmetric nuclear 
matter, in particular the relation between the nature of fluctuations,
the types of instabilities and the properties of the interaction. We show
a chemical instability appears as an instability against isoscalar-like
fluctuations. Then starting from 
phenomenological hadronic field theory (QHD), including  exchange terms,
we discuss the symmetry energy and the relation to the dynamical
response inside the spinodal region.}

\section{Introduction}
Nuclear matter (NM) is a binary system (neutrons and protons),
this feature has stimulated a quite exciting research field: nuclei far from
$\beta$-stability, dynamical effects of a large charge asymmetry, properties
of neutron stars. In particular with radioactive beams (not only), NM
with a high isospin asymmetry can be created transiently, offering the 
possibility to study also chemical instability associated with isospin
asymmetry of NM. Indeed, recent experiments on nuclear reactions 
involving different ratios of the neutron to proton numbers have
shown that products from nuclear multifragmentation depends strongly
on the isospin asymmetry of the colliding nuclei \cite{Sob97,Prak97}.\par
In the first part of the paper we will focus on the behaviour 
of Asymmetric NM (ANM) at sub-normal
densities where we can foresee scenarios for a dynamical formation of 
fragments with particular isotopic contents due to instabilities.
To study these problems we show, by linking thermodynamics and 
Fermi liquid theory, the relation between the nature of fluctuations and 
the types of instabilities for a general binary system.
In the second part we show a relativistic field model of hadrons (QHD)
based on strongly interacting nucleon and meson fields.
Inside the QHD model we have recently proposed a method to introduce
Fock term contributions \cite {prc_nuovo}.
Here we want to stress some implications of that
improvement on the isospin physics discussed in the first part.
Comparisons with the QHD model in Hartree approximation and with 
Skyrme interactions will be addressed.
    
\section{Chemical and mechanical instabilities}
\subsection{Analytical analysis}
In the context of multifragmentation the instability of NM (region where the
system becomes unstable against long wave length but small amplitude
fluctuation) plays a crucial role. 
These aspects were discussed repeatedly in the past,
\cite{Peth88,Peth93,col94,cdg94,col98}
but the binary character will induce new features for both scenarios that are
absent in one-component systems.
  In the framework of Landau theory for two component Fermi liquids the
spinodal border was determined by studying the stability of collective modes
described by two coupled Landau-Vlasov equations for protons and neutrons.
In terms of the appropriate Landau parameters the stability condition can be
expressed as~\cite{bar98},
\begin{equation}
(1 + F_0^{nn})(1 + F_0^{pp}) - F_0^{np}F_0^{pn} > 0~.
\label{eq:Land}
\end{equation}
It is possible to show that this condition is equivalent to
the following thermodynamical condition \cite{bar00}
\begin{equation}
\left({\partial P \over \partial \rho}\right)_{T,y}
\left({\partial\mu_p \over \partial y}\right)_{T,P} > 0~.
\label{eq:ser}
\end{equation}
discussed in \cite{landau,serot}, where $y$ is the proton fraction.
In Fig. 1 we show the spinodal lines obtained from eq.~(\ref{eq:Land})
(continuous line with dots) which for asymmetric nuclear matter is seen to
contain the lines corresponding to
"mechanical instability",
$\left({\partial P \over \partial \rho}\right)_{T,y}<0$
(crosses). Therefore both eqs. (\ref{eq:Land},\ref{eq:ser}) describe the 
whole region of instability.
\begin{figure}[htb]
\epsfysize=3.8cm
\centerline{\epsfbox{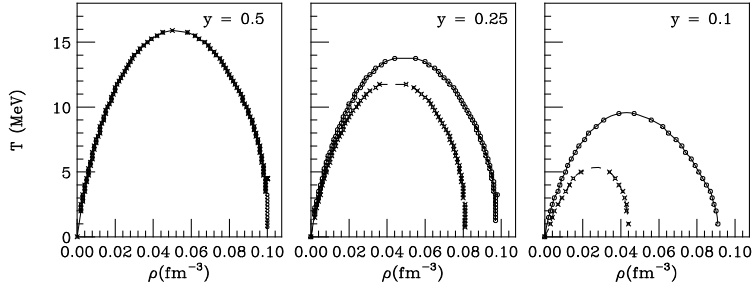}}
\caption
{Spinodal lines corresponding to chemical (circles) and
mechanical (crosses) instability for three value of proton fraction $y$.}
\end{figure}
We want to stress, however, that by just looking at the above stability 
condition
we cannot determine the nature of the fluctuations against which a binary
system becomes chemically unstable. Indeed, the thermodynamical
condition in eq.~(\ref{eq:ser}) cannot distinguish between two  very different
situations which can be encountered in nature: an attractive
interaction between the two components of the mixture ($F_0^{np},F_0^{pn} <
0$),
 as is the case of nuclear matter, or a  repulsive interaction
between the two species. \par
We define as isoscalar-like density fluctuations
the case when proton and neutron Fermi spheres
(or equivalently the proton and neutron densities) fluctuate in phase and as
isovector-like density fluctuations when the two Fermi sphere fluctuate
out of phase. Then it is possible to prove \cite{bar00}, based on
a thermodynamical approach of asymmetric Fermi liquid mixtures,that
chemical instabilities are triggered by isoscalar fluctuations in the first,
i.e. attractive, situation
and by isovector fluctuations in the second one. For the asymmetric nuclear
matter case because of the attractive interaction between protons and neutrons
the phase transition is thus due to isoscalar fluctuations that induce
chemical instabilities while the system is never unstable against isovector
fluctuations.
Of course the same attractive interaction is also at the origin of phase
transitions in symmetric nuclear matter. However, in the asymmetric case
isoscalar fluctuations lead to a more symmetric high density phase
everywhere under the instability line defined by 
eq.(\ref{eq:Land})~\cite{bar98}
\begin{figure}[htb]
\epsfysize=4.5cm
\centerline{\epsfbox{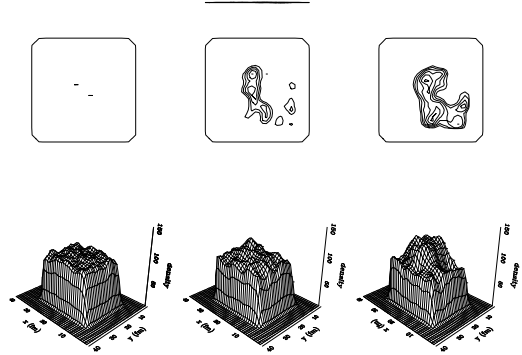}}
\caption
{ Time evolution of the density $\rho(x,y)$ in the plane $z$=0
for initial density $\rho^{(0)}=0.09 fm^{-3}$, at T=5 MeV and asymmetry
$I=0.5$. Upper panels show contour plots of $\rho(x,y)$  and lower panels
the corresponding two dimensional surface.}
\end{figure}
\subsection{Numerical results: heated nuclear matter in a box}
The previous discussion is based on a link between thermodynamics
and Fermi liquid approaches. Then numerical approach were performed in
order to follow all stages of fluctuation development in
the fragment formation process. In the numerical 
approach we consider nuclear matter in a box of size L=24fm
 imposing periodic boundary conditions.\par
We follow a phase space test particle method to solve 
the Landau-Vlasov dynamics (using Gaussian wave packets) so the dynamics of
nucleon-nucleon collision is also included.
An initial temperature is introduced by distributing the test particle
momenta according to Fermi distribution. We have followed the space-time
evolution of test particles for a value of the
initial asymmetry I=0.5, at initial density $\rho^{(0)}=0.09fm^{-3}$ and T=5 MeV
in such way we start from the region of chemical instability (see Fig.1).
The initial density perturbation was created automatically due to the random
choice of test particle positions.
We report in Fig.2 density distribution in the plane $z$=0 at three time steps
$t$=0,100,200 fm/c,corresponding respectively to initial
conditions, intermediate and final stages of the spinodal decomposition (SD).
The contour plots delimit the region with density higher than the initial value
of density.\par
\begin{figure}[htb]
\epsfysize=4.5cm
\centerline{\epsfbox{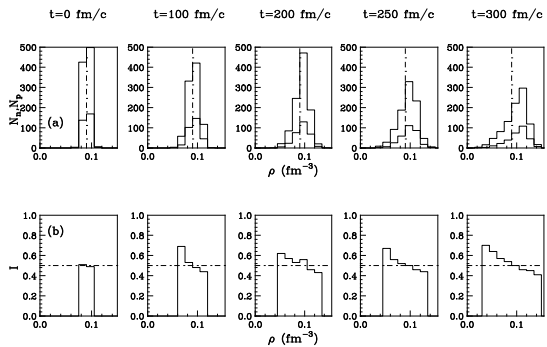}}
\caption
{Time evolution of neutron (thicks lines) and proton (thin lines)
abundance (a) and asymmetry (b) as function of density.}
\end{figure}
We report the time evolution of neutron (thick histogram) and proton
(thin histogram) abundance (Fig.3a) and asymmetry in various density 
bins (Fig.3b).
The dashed line respectively shows the initial uniform density value 
$\rho^{(0)}= 0.09 fm^{-3}$ (3a) 
and the initial asymmetry I=0.5 (3b). The drive to 
higher density regions is clearly different for neutrons and protons:
at the end of the dynamical clustering mechanism we have different asymmetries 
in the liquid and gas phases (see the panel at 200 fm/c).
This result is the same of ref.10 where the dynamics of 
mechanical instabilities 
were studied, demonstrating that also in a complete dynamical calculation
the kind of fluctuation associated with chemical and mechanical instabilities
are the same.
\section{Quantum hadrodynamics model (QHD)}
\subsection{Treatment of Fock terms}
Phenomenological hadronic field theories (Quantum Hadrodynamics, $QHD$)
are widely used in dense nuclear matter studies \cite{sewa86}.
In most of the previous works on the subject, the Relativistic Mean
Field ($RMF$) approximation of $QHD$ has been followed. In the $RMF$
the meson fields are treated as classical fields and
a Hartree reduction of one body density matrices is used.
This implies that
each meson field is introduced, with appropriated readjusted couplings,
just to describe the dynamics of a corresponding degree of freedom,
without mixing due to many-body effects: neutral $\sigma$ and $\omega$
mesons are in charge of saturation properties, isospin effects are carried
by isovector $\delta$ and $\rho$ mesons. In a sense the model represents
a straightforward extension of the One-Boson-Exchange ($OBE$) description of
nucleon-nucleon scattering.\par
Our aim is to introduce explicit many-body effects
just evaluating exchange term contributions. 
Fock terms play an essential role in symmetry breaking and consequent mixing
of different degrees of freedom.
In particular, in the context of the QHD model, essential properties
of nuclear matter come mostly from the two neutral strong meson fields. Hence
it is important to evaluate the Fock contribution associated  with these
fields. \par
Then we will start from a {\it $QHD-II$} model \cite{sewa86}
where the nucleons
are coupled to neutral scalar $\sigma$ and vector $\omega$ mesons
and to the isovector $\rho$ meson.
Self-interaction terms of the $\sigma$-field were
originally introduced for renormalization reasons \cite{bobo77,bopr89}
and can also be considered as a way to parametrize the density dependence
of $NN$ force. Actually they are also describing medium effects essential 
to reproduce important properties
(compressibility and nucleon effective mass) of nuclear matter
around saturation density.

The Lagrangian density for this model is given by:
\begin{eqnarray}
{\cal L} = {\bar {\psi}}[\gamma_\mu(i{\partial^\mu}-{g_V}{\cal V}^\mu
 - g_{\rho}{\bf {\cal B}}^\mu \cdot {\bf{\tau}} )-
(M-{g_S}\phi)]\psi + {1 \over 2}({\partial_\mu}\phi{\partial^\mu}\phi
- {m_S}^2 \phi^2) \nonumber \\
- {a \over 3} \phi^3 - {b \over 4} \phi^4
- {1 \over 4} W_{\mu\nu}
W^{\mu\nu} + {1 \over 2} {m_V}^2 {\cal V}_\nu {{\cal V}^\nu}
- {1 \over 4} {\bf L}_{\mu\nu}\cdot
{\bf L}^{\mu\nu} + {1 \over 2} {m_{\rho}}^2 {\bf {\cal B}}_\nu\cdot
{{\bf {\cal B}}^\nu}
\end{eqnarray}
where
$W^{\mu\nu}(x)={\partial^\mu}{{\cal V}^\nu}(x)-
{\partial^\nu}{{\cal V}^\mu}(x)~$ and
${\bf L}^{\mu\nu}(x)={\partial^\mu}{{\bf {\cal B}}^\nu}(x)-
{\partial^\nu}{{\bf {\cal B}}^\mu}(x)~.$
Here $\psi(x)$ generally denotes the
fermionic field, $\phi(x)$ and ${{\cal V}^\nu}(x)$ represent neutral scalar and
a vector boson fields, respectively. ${{\bf {\cal B}}^\nu}(x)$
is the charged vector
field and ${\bf{\tau}}$ denotes the isospin matrices.\par
We have treated Hartree-Fock (HF) terms  in EOS and transport
equation at the same level, for this reason we performed
the many-body calculations in the
quantum phase space introducing the Wigner transform of the
one-body density matrix of the fermion field. This method has
the advantages of a direct derivation of dynamical transport equations 
\cite{mat91}.
In ref. \cite{prc_nuovo} the way we introduce Fock terms is discussed in
detail, here we want to discuss some results for $EOS$ of ANM.
\begin{figure}[htb]
\epsfysize=4.5cm                  
\centerline{\epsfbox{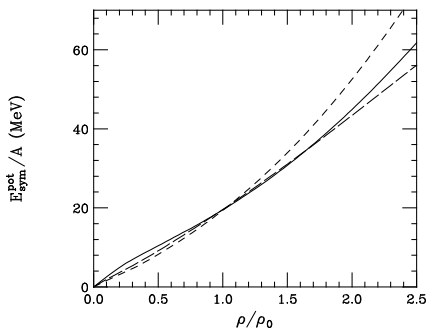}}
\caption
{Symmetry energy per nucleon vs. baryon density.
Long dashed line: $NLH$ with
$\rho$ meson. Solid line: $NLHF$ results.
Dashed line: $NLH$ with $\rho$ and $\delta$ mesons.}
\end{figure}
\subsection{Asymmetric Nuclear Matter}
We obtain, thanks to Fock terms, scalar and vector isovector contributions 
to symmetry energy, generally associated respectively with $\delta$ and
 $\rho$ mesons, even without isovector mesons \cite{gre01}.
We show the
comparison between our Non Linear Hartree-Fock (NLHF) and those of the
Non linear Hartree (NLH), including the isovector $\rho$ and $\delta$
mesons \cite{kuku97}, with parameters fitted in order to give the same 
saturation properties and a symmetry energy at $\rho_0$ of 31 MeV.
The region in (T,$\rho$) plane at different asymmetry of chemical and 
mechanical region are quite similar \cite{liu01} to the one of Skyrme 
interaction (see Fig.1).
 
In Fig.4 the symmetry energy  obtained in NLHF
(solid line) in comparison with the result of NLH including both
$\rho$ and $\delta$ meson (dashed line). 
Actually the kinetic contribution is subtracted (anyway it is the same in all
the models),so we will refer to potential symmetry energy $E_{sym}^{pot}$.
For reference
the most common result,  including only $\rho$ as isovector meson,
is plotted (long dashed line).  
In all these relativistic models a quite repulsive density dependence 
of the simmetry term is obtained,
but we notice that the density dependence of $E_{sym}^{pot}$ in the complete 
NLH+$\rho$+$\delta$ model is quite different respect to our result. This is
due to the fact that in NLHF the coupling in the isovector channels 
become density dependent. This implies a "softer" behaviour of the potential
symmetry term {\it below the saturation density} in the NLHF case
and a "stiff" behaviour more similar to NLH +$\rho$ above.\par
\begin{figure}[htb]
\epsfysize=4.5cm
\centerline{\epsfbox{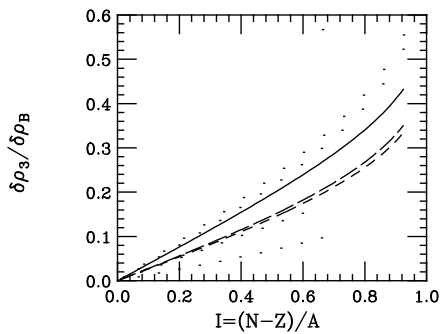}}
\caption
{Ratio of isovector and isoscalar amplitudes as function of
asymmetry I at half saturation density. The lines have the same
meaning of Fig.4}
\end{figure}
A transport equation can
be consistently derived to be used for the study of dynamical evolution of
nuclear matter far from normal conditions. We expect that the dynamical
evolution can point out the important difference between NLH and NLHF 
by comparison
with experimental data for isospin of IMF (in multifragmentation or
neck events) and for isospin flows. We have just found as the latter are
affected by density behaviour of $E_{sym}^{pot}$ and the nucleon effective
mass splitting \cite{dtr00}.
In this view ve have performed the study of collective response in ANM
by means of a relativistic kinetic equation in linear approximation
respect to the fluctuation \cite{ma94}.
In Fig.5 we present the unstable isoscalar-like solution of the dispersion
relation for low density NM (the density is $\rho=0.4 \rho_0$), 
in particular the ratio ${\delta\rho_3/ \delta\rho_b}$ as function
of the initial asymmetry.
We notice that the NLHF tendency to restore the isotopical symmetry in the 
spinodal decomposition of neutron rich NM is different respect to NLH result 
(both dashed and long dashed result). In particular we stress that the 
discrepancy is equal to the one obtained \cite{cdl98} with Skyrme interaction, 
in a non relativistic approach, comparing a Asy-soft with an Asy-stiff term
for symmetry energy.
However at variance with Skyrme results the behaviour of $E_{sym}$
in NLHF appear to be Asy-soft at subnuclear density but it's Asy-stiff above 
saturation density. A mixed behaviour is predicted with expected interesting
effects on experimental observable depending on the probed barion
density region of the interacting nuclear matter.

\end{document}